\documentclass[reprint, superscriptaddress, showpacs]{revtex4-1}
\usepackage{graphicx}
\usepackage{caption}
\usepackage{subfig}
\usepackage{float}
\usepackage{caption}
\usepackage{ragged2e}
\usepackage{amsfonts}
\usepackage{amsmath}

\begin{document}


\title{Pygmy resonance and low-energy enhancement in the $\gamma$-ray strength functions of Pd~isotopes}

\author{T.~K.~Eriksen}
\email{t.k.eriksen@fys.uio.no}
\affiliation{Department of Physics, University of Oslo, N-0316 Oslo, Norway}
\author{H.~T.~Nyhus}
\affiliation{Department of Physics, University of Oslo, N-0316 Oslo, Norway}
\author{M.~Guttormsen}
\affiliation{Department of Physics, University of Oslo, N-0316 Oslo, Norway}
\author{A.~G\"{o}rgen}
\affiliation{Department of Physics, University of Oslo, N-0316 Oslo, Norway}
\author{A.~C.~Larsen}
\affiliation{Department of Physics, University of Oslo, N-0316 Oslo, Norway}
\author{T.~Renstr\o m}
\affiliation{Department of Physics, University of Oslo, N-0316 Oslo, Norway}
\author{I.~E.~Ruud}
\affiliation{Department of Physics, University of Oslo, N-0316 Oslo, Norway}
\author{S.~Siem}
\affiliation{Department of Physics, University of Oslo, N-0316 Oslo, Norway}
\author{H.~K.~Toft}
\affiliation{Department of Physics, University of Oslo, N-0316 Oslo, Norway}
\author{G.~M.~Tveten}
\affiliation{Department of Physics, University of Oslo, N-0316 Oslo, Norway}
\author{J.~N.~Wilson}
\affiliation{Institut de Physique Nucl\'eaire, 91406 - Orsay Cedex, France}

\begin{abstract}
\begin{description}
\item[Background]
An unexpected enhancement in the $\gamma$-ray strength function, as compared to the low energy tail of the Giant Dipole Resonance (GDR), has been observed for Sc, Ti, V, Fe and Mo isotopes for $E_\gamma<4$~MeV. This enhancement was not observed in subsequent analyses on Sn isotopes, but a Pygmy Dipole Resonance (PDR) centered at $E_\gamma\approx8$~MeV was however detected. The $\gamma$-ray strength functions measured for Cd isotopes exhibit both features over the range of isotopes, with the low-energy enhancement decreasing- and PDR strength increasing as a function of neutron number. This suggests a transitional region for the onset of low-energy enhancement, and also that the PDR strength depends on the number of neutrons.
\item[Purpose]
The $\gamma$-ray strength functions of $^{105-108}$Pd have been measured in order to further explore the proposed transitional region. 
\item[Method] 
Experimental data were obtained at the Oslo Cyclotron Laboratory by using the charged particle reactions ($^{3}$He, $^{3}$He$^{\prime}\gamma$) and ($^{3}$He, $\alpha$$\gamma$) on $^{106,108}$Pd target foils. Particle$-\gamma$ coincidence measurements provided information on initial excitation energies and the corresponding $\gamma$-ray spectra, which were used to extract the level densities and $\gamma$-ray strength functions according to the Oslo method.
\item[Results]
The $\gamma$-ray strength functions indicate a sudden increase in magnitude for $E_{\gamma}>4$~MeV, which is interpreted as a PDR centered at $E_{\gamma}\approx8$~MeV. An enhanced $\gamma$-ray strength at low energies is also observed for $^{105}$Pd, which is the lightest isotope measured in this work.
\item[Conclusions]
A PDR is clearly identified in the $\gamma$-ray strength functions of $^{105-108}$Pd, and a low-energy enhancement is observed for $^{105}$Pd.
Further, the results correspond and agree very well with the observations from the Cd isotopes, and support the suggested transitional region for the onset of low-energy enhancement with decreasing mass number. The neutron number dependency of the PDR strength is also evident.
\end{description}
\end{abstract}


\maketitle 

\section{Introduction}
\label{sec:intro}
Astrophysical models aiming at explaining the nature of the s- and r-process nucleosyntheses are highly dependent on neutron capture cross sections and corresponding reaction rates. This is also true for more applied cases, e.g.~modeling of isotope production in reactors. The $\gamma$-ray strength function is an important input parameter in calculations of radiative neutron capture (n,$\gamma$) cross sections, and information on the $\gamma$-ray strength function for energies below the neutron separation energy is essential for reliable estimates of these cross sections.

Nuclear level densities and $\gamma$-ray strength functions are average quantities used to describe nuclear thermodynamic and electromagnetic properties, respectively, in the quasi-continuum of excited states.
The onset of quasi-continuum is typically at a few MeV of excitation energy above the ground state, and denotes the region of energy where the density of levels is so high that their widths and level spacing are comparable in size.
The nuclear physics group at the University of Oslo has developed the Oslo method, which allows for extraction of both level density and $\gamma$-ray strength from the onset of quasi-continuum and up to the nucleon binding energies~\cite{extraction_NLD_gammastr}.
The present work concerns analyses of these quantities for $^{105-108}$Pd, with most focus on the $\gamma$-ray strength functions.

In previous analyses of $^{43-45}$Sc~\cite{sc_upb1, sc_upb2}, $^{44-46}$Ti~\cite{ti_upb1, ti_upb2, ti_upb3}, $^{50,51}$V~\cite{v_upb1}, $^{56,57}$Fe~\cite{iron_upb2} and $^{93-98}$Mo~\cite{moly_upb1} isotopes using the Oslo method, an unexpected enhancement in the $\gamma$-ray strength was discovered at low $\gamma$-energies, i.e.~$E_\gamma<4$~MeV. This low-energy enhancement was recently supported by results from a different experimental approach for $^{95}$Mo~\cite{moly_upb2}, which gives confidence to the results of the Oslo method. The feature has drawn a lot of attention, and it has recently been shown that the low-energy enhancement in $^{56}$Fe is dominated by dipole transitions~\cite{56Fe_ang}. However, the electromagnetic character has not yet been determined, and there are theoretical explanations suggesting both electric-~\cite{lowE_E1} and magnetic~\cite{lowE_M1} characters.

In similar analyses of $^{116-119}$Sn~\cite{tin_pyg1} and $^{121,122}$Sn~\cite{tin_pyg2} using the Oslo method, there were no signs of the low-energy enhancement. However, enhancement at higher energies ($E_\gamma>4$~MeV) was observed for these nuclei, and this was interpreted as a Pygmy Dipole Resonance (PDR) centered at $E_\gamma\approx8$~MeV. 

The motivation for investigating the Pd isotopes was to further examine the $\gamma$-ray strength functions for nuclei in the mass region where the characteristics of the $\gamma$-ray strength function seem to change. Indications of a transition have recently been observed for $^{105,106,111,112}$Cd~\cite{Cd-isotopes}, where the results show enhancement at low energy ($E_\gamma<4$~MeV) for $^{105,106}$Cd, but not for $^{111,112}$Cd. Enhanced strength for $E_\gamma>4$~MeV is observed for all the Cd isotopes, corresponding to the PDR seen in Sn isotopes. However, the Cd isotopes show that the PDR strength increases as a function of neutron number, which was not seen for the Sn isotopes. The Pd isotopes investigated in this work are very close to the Cd isotopes in both proton- and neutron numbers, and the results are expected to reveal more information on these matters.

The article is structured in the following way: The experimental approach and the Oslo method are explained in Secs. \ref{sec:experiment} and \ref{sec:extraction}. Analyses and results are discussed in Secs. \ref{sec:ld} and \ref{sec:g-rsf}, and a concluding summary is provided in Sec. \ref{sec:summary}.

\section{Experiment}
\label{sec:experiment}
The experiments were conducted at the Oslo Cyclotron Laboratory (OCL) at the University of Oslo, where the MC-35 Scanditronix cyclotron was used to accelerate $^{3}$He ions to a kinetic energy of 38~MeV. In two separate runs, the accelerated ion-beam was directed at self supporting $^{106,108}$Pd target foils of thicknesses 1~mg/cm$^2$, and the excited states of $^{105-108}$Pd were populated through the charged-particle reactions ($^{3}$He, $\alpha$$\gamma$) and ($^{3}$He, $^{3}$He$^{\prime}\gamma$). The energies of the ejected particles and coinciding $\gamma$-ray emissions were measured for a period of seven days in both runs, and detected events were stored in list mode for offline sorting. 

Particle energies were measured with SiRi~\cite{NIM}, which is a composite detector system consisting of 8 trapezoidal-shaped silicon $\Delta E-E$ telescopes put together to form a hollow, truncated cone-like geometry. The modules consist of a 1550-$\mu$m thick $E$ detector with a 130-$\mu$m thick $\Delta E$ detector in front, and the $\Delta E$ detectors are further segmented into 8 curved strips covering scattering angles between 40$^{\circ}$ and 54$^{\circ}$ relative to the beam direction. This makes up 64 particle telescopes in total. The system was positioned in forward direction, with the center of the detector modules at an angle of 45$^{\circ}$ and a distance of $5.0$~cm from the target.

Coincident $\gamma$-rays were measured by CACTUS, a detector system consisting of 28 spherically distributed, collimated, 5~$^{\prime\prime}\times$5~$^{\prime\prime}$ NaI(Tl) $\gamma$-ray detectors. The detectors have a total efficiency of $\approx15$\% of 4$\pi$, and an energy resolution of $\approx7$\% FWHM at $E_{\gamma}=1332$~keV. The detector front ends were positioned $22.0$~cm from the center of the target. 

The measured events were sorted according to reaction channels by gating on the corresponding $\Delta E-E$ curves, and the resulting particle- and  $\gamma$-ray energy spectra were calibrated to known level- and $\gamma$-transition energies. The excitation energy of residual nuclei was calculated from the reaction kinematics, and the measured data were arranged in ($E_{\gamma}$,$E_x$) coincidence matrices, where $E_{\gamma}$ and $E_x$ are the $\gamma$-ray and excitation energies, respectively. The raw coincidence matrix of $^{107}$Pd is depicted in Fig.~\ref{fig:coincidence_matrices}a).

\section{The Oslo method}
\label{sec:extraction}
The first step of the Oslo method is to unfold the raw $\gamma$-ray spectra, which has to be done before useful information can be extracted from the coincidence matrices. The $\gamma$-ray spectra were unfolded by applying the folding iteration method~\cite{unfolding} with the measured response functions of CACTUS. This procedure corrects the $\gamma$-ray spectra for unwanted contributions due to the detector response.

\begin{figure}
\includegraphics[width=\columnwidth]{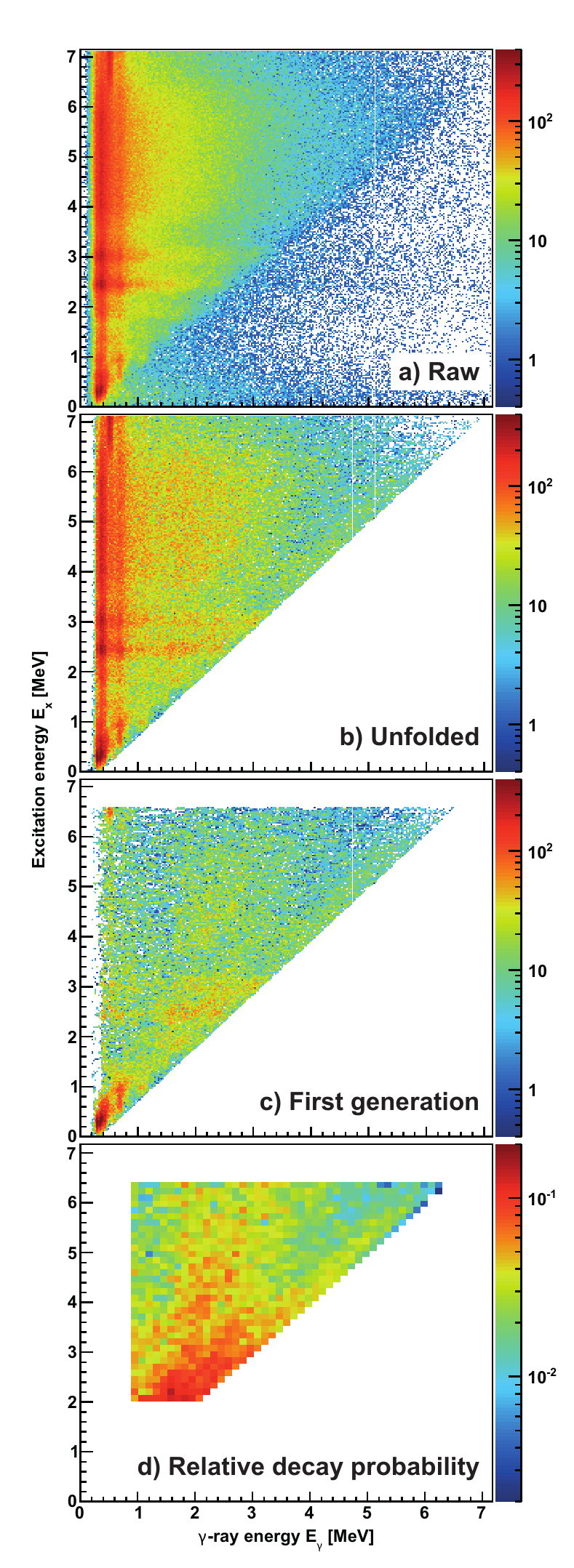}
\caption[Coincidence matrices]{(Color online) Coincidence matrices for $^{107}$Pd. Details are provided in the text.}
\label{fig:coincidence_matrices}
\end{figure}

Further, primary $\gamma$-rays have to be deduced from the unfolded spectra.
Extraction of primary $\gamma$-rays is necessary because $\gamma$-decay may, and generally does, occur through a cascade of transitions that cannot be distinguished in time. As a consequence, the measured $\gamma$-ray spectra contain all generations of $\gamma$-rays in the cascade, but only the first generation (i.e.~primary) $\gamma$-rays provide information which is relevant to the Oslo method. A method for extracting the first generation $\gamma$-ray spectra has been developed~\cite{firstgenmatrix}, in which a weighted sum of all $\gamma$-spectra corresponding to $E_x<E_x^\prime$ is subtracted for each $E_x^\prime$. The weights are found by an iterative procedure, and matrices of first-generation $\gamma$-ray spectra are extracted from the unfolded ($E_{\gamma}$,$E_x$) coincidence matrices. Figures \ref{fig:coincidence_matrices}a) - c) show the matrices of $^{107}$Pd for each step of this procedure.

The $\gamma$-ray spectra of the first-generation matrix are then normalized to unity for each $E_x$ bin. This is performed for energies above $E_{\gamma,\rm min}$ and $E_{x,\rm min}$, and results in a matrix of relative $\gamma$-decay probabilities $P(E_{\gamma},E_x)$. The lower limit $E_{\gamma,\rm min}$ is determined on basis of the first-generation $\gamma$-ray spectra, because the extraction method generally leads to over-subtraction at low $\gamma$-ray energies due to a mismatch with the spin distribution in the lowest excitation energy region. The lower limit $E_{x,\rm min}$ is set to exclude discrete levels from the analysis. Figure~\ref{fig:coincidence_matrices}d) shows the normalized matrix for $^{107}$Pd.

The probability of $\gamma$-decay from an initial state $E_x$ to a final state $E_f$ by a $\gamma$-ray of energy $E_{\gamma}=E_x-E_f$, is proportional to the level density at the final state $\rho(E_f)$ and a $\gamma$-ray energy dependent transmission coefficient $\mathcal{T}(E_{\gamma})$. 
Hence, the normalized first generation $\gamma$-ray matrix can be factorized into~\cite{extraction_NLD_gammastr}
\begin{equation}
\label{eq:factor}
P(E_{\gamma},E_x)\propto \mathcal{T}(E_{\gamma}) \rho(E_x-E_{\gamma})~,
\end{equation}
which is built on the assumption that the nucleus reaches a compound state after excitation, and that the manner of the subsequent $\gamma$-decay is mainly statistical and independent on how the state was formed. According to the Brink hypothesis~\cite{brink}, any collective decay mode has the same properties whether it is built on the ground state or on an excited state, and the $\gamma$-ray transmission coefficient is therefore assumed to depend on $\gamma$-ray energy only. It can also be noted that the factorization is closely related to Fermi's golden rule, e.g.~Ref.~\cite{krane}.

A mathematical representation of the relative $\gamma$-decay probability matrix is given by the expression~\cite{extraction_NLD_gammastr}
\begin{equation}
\label{eq:fg_theo}
P_{\rm th} (E_{\gamma},E_x)=\frac{\mathcal{T}(E_{\gamma}) \rho(E_x-E_{\gamma})}{\sum_{E_{\gamma}=E_{\gamma, \rm min}}^{E_x} \mathcal{T}(E_{\gamma}) \rho(E_x-E_{\gamma})}~,
\end{equation}
and unique functional forms of $\mathcal{T}(E_{\gamma})$ and  $\rho(E_x-E_{\gamma})$ are derived by fitting Eq.~(\ref{eq:fg_theo}) to the matrices of relative decay probabilities by a least squares method described in Ref.~\cite{extraction_NLD_gammastr}.  
Unfortunately, there is an infinite set of equally good normalizations for the two extracted functions that lead to reproduction of $P_{\rm th}(E_{\gamma},E_x)$. However, all the solutions can be reached by applying the transformations~\cite{extraction_NLD_gammastr}
\begin{equation}
\label{eq:trans_1}
\tilde{\rho}(E_x-E_{\gamma})=\rho(E_x-E_{\gamma}) A e^{\alpha (E_x-E_{\gamma})}~,
\end{equation}
\begin{equation}
\label{eq:trans_2}
\tilde{\mathcal{T}}(E_{\gamma})=\mathcal{T}(E_{\gamma}) B e^{\alpha E_{\gamma}}~,
\end{equation}
where $A$ and $B$ are scaling coefficients, and $\alpha$ adjusts the slopes of the functions.
In order to determine the most physical solutions of Eqs.~(\ref{eq:trans_1}) and (\ref{eq:trans_2}), the extracted data are normalized to known experimental data as described in the following. 

The determination of $A$ and $\alpha$ is performed by normalizing the extracted level density at both low and high excitation energies. At low excitation energies, this is done by matching the extracted level density to the number of known levels per $E_x$ bin. In the high energy region, it is normalized to a semi-experimental level density derived from the Back Shifted Fermi Gas (BSFG) model and data from neutron resonance experiments. 

In the BSFG model, the total level density for all spins and parities is given by~\cite{nld_parameters}
\begin{equation}
\label{eq:leveldensity_all}
\rho(E_x)=\frac{1}{12 \sqrt{2} \sigma} \frac{e^{2\sqrt{a (E_x-E_1)}}}{a^{1/4} (E_x-E_1)^{5/4}}~,
\end{equation}
where $E_x$ is the excitation energy, $a$ is the level density parameter, $E_1$ is the energy backshift parameter, and $\sigma$ is the spin-cutoff parameter. 
Further, the spin dependent level density is described by
\begin{equation}
\label{eq:leveldensity_spindep}
\rho(E_x, J)=\rho(E_x) \left[ \frac{(2J+1) e^{-(J+1/2)^2/2\sigma^2}}{2 \sigma^2} \right]~,
\end{equation}
where $J$ denotes the spin of the nucleus. The braced part of Eq.~(\ref{eq:leveldensity_spindep}) is the spin distribution $g(E_x,J)$ of the level density~\cite{GetC}, 
and the spin-cutoff parameter is given by~\cite{nld_parameters}
\begin{equation}
\label{eq:spin-cutoff}
\sigma^2(E_x)=0.391 A^{0.675} (E_x-0.5Pa')^{0.312}~,
\end{equation} 
where $Pa'$ is the deuteron pairing energy. 
The uncertainty of the spin-cutoff parameter was determined by assuming that the lowest reasonable value is 10\% less than calculated by Eq.~(\ref{eq:spin-cutoff}), and that the highest reasonable value is 5\% higher than estimated by~\cite{nld_parameters_2006}
\begin{equation}
\label{eq:spin-cutoff_old}
\sigma^2(E_x)=0.0146 A^{5/3} \frac{1+\sqrt{1+4a(E_x-E_1)}}{2a}~.
\end{equation} 
This approach was chosen because Eq.~(\ref{eq:spin-cutoff_old}) gives a relatively higher value of the spin-cutoff parameter than Eq.~(\ref{eq:spin-cutoff}).

In a neutron resonance experiment where $I_t$ is the spin of the target nucleus, and when assuming equal parity distribution, the neutron resonance spacing $D_{0}$ can be written in terms of the spin dependent level density as
\begin{equation}
\label{eq:n_resonance_spacing}
\frac{1}{D_{0}}=\frac{1}{2} \sum_j \rho(B_n, |I_t\pm j|)~,
\end{equation}
where $j=|\ell\pm s|$ represents the component of the total angular momentum of the neutron. Since $D_0$ denotes the resonance spacing for s-wave neutrons, it implies that $\ell=0$ and hence $j=1/2$. The relation of Eq.~(\ref{eq:n_resonance_spacing}) is justified by the fact that all levels with $J_f=|I_t\pm 1/2|$ is accessible in an s-wave neutron resonance experiment, and the expression is divided by 2 due to the assumption of equal parity distribution at the neutron binding energy. The total $\rho(B_n)$ is found by combining Eqs.~(\ref{eq:leveldensity_spindep}) and (\ref{eq:n_resonance_spacing}), and rearranging with respect to the level density,
\begin{equation}
\label{eq:leveldensity_bn}
\rho(B_n)=\frac{2}{D_{\ell}} \frac{1}{\sum_j g(B_n,J_f)}.
\end{equation}

\begin{figure}
\includegraphics[width=\columnwidth]{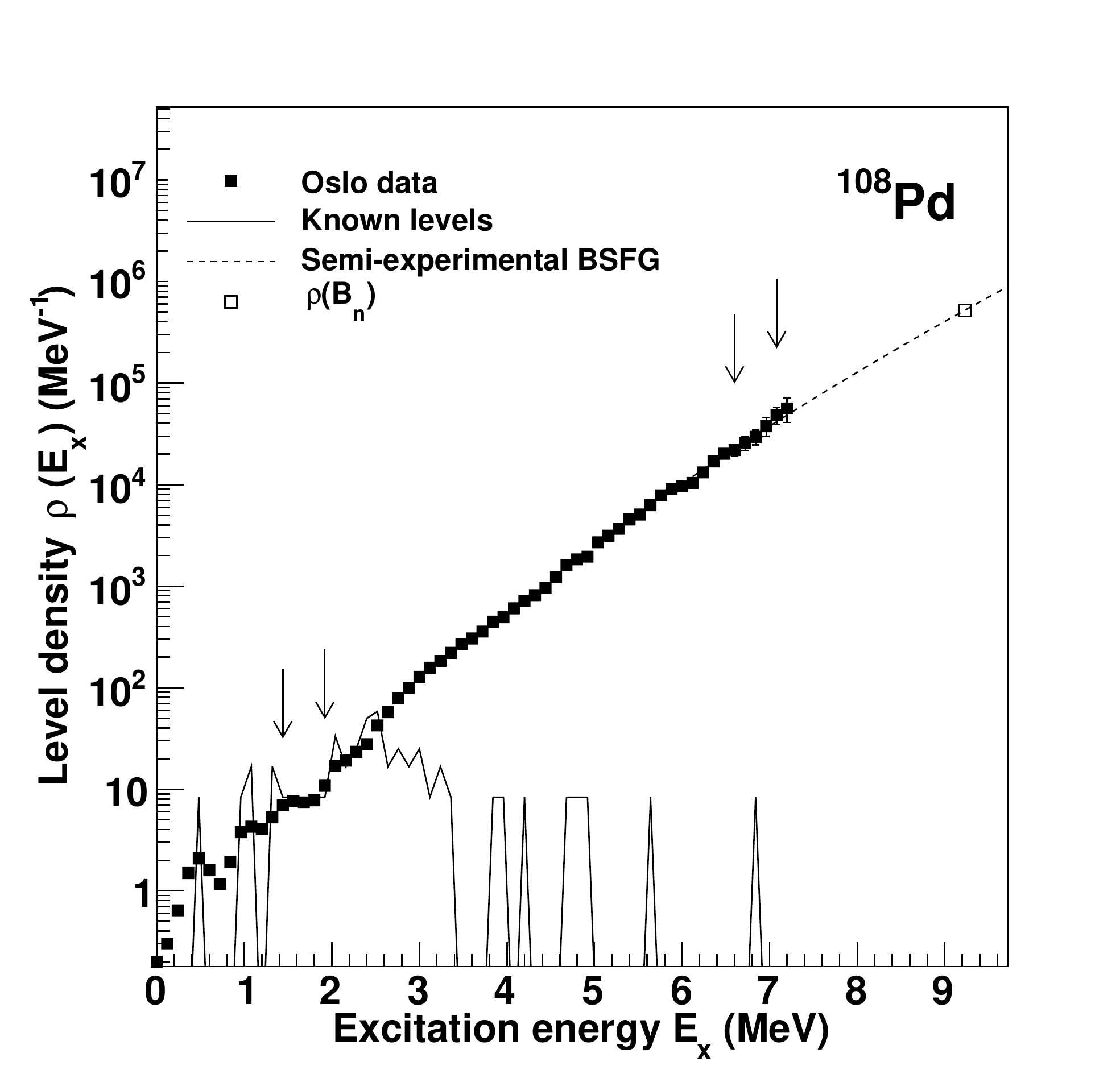}
\caption[LD norm 108Pd]{Normalization of the level density of $^{108}$Pd. Data points are fitted between the arrows, and further details are provided in the text.}
\label{fig:ld_norm}
\end{figure}

The semi-experimental level density, to which the experimental data are normalized at high $E_x$, is given by Eq.~(\ref{eq:leveldensity_all}) and scaled to match the value of the deduced $\rho(B_n)$. Interpolation by this semi-experimental level density is necessary because the experimental data can only be extracted up to $E_x=B_n-E_{\gamma,{\rm min}}$. Figure~\ref{fig:ld_norm} depicts the normalization for $^{108}$Pd.

The absolute normalization of $\mathcal{T}(E_{\gamma})$, i.e.~finding the scaling parameter $B$, is performed by using experimental values of the average total radiative width $\left< \Gamma_{\gamma}\right>$ at the neutron binding energy, and the s-wave neutron resonance spacing $D_{0}$.
The average total radiative width of excited states with energy $E_x$, spin $J$ and parity $\pi$ can be described by~\cite{atrw_E}
\begin{align}
\label{eq:atrw_E}
&\left< \Gamma_{\gamma}(E_x,J,\pi)\right>=\nonumber\\ 
&~~~~~\frac{1}{2\pi \rho(E_x,J,\pi)}\sum_{X L}{}\sum_{J^{\prime},\pi^{\prime}}{}\int_{E_{\gamma}=0}^{E_x}dE_{\gamma} \mathcal{T}_{X L}(E_{\gamma})\nonumber\\ 
&~~~~~~~~~~~~~~~~~~~~~~~~~~~~~~~~~~~\times \rho(E_x-E_{\gamma},J^{\prime},\pi^{\prime})~,
\end{align}
where $X$ and $L$ denotes the electromagnetic character and multipolarity respectively, and the summation and integration are over all final states with 
\begin{equation}
J^{\prime}=\sum^{L}_{L^{\prime}=-L} J+L^{\prime},
\end{equation} 
and $\pi^{\prime}$ accessible by $\gamma$-transitions of energy $E_{\gamma}$.

It is well known that nature favors the lowest multipolarity allowed for a transition, and due to the high density of levels in the quasi-continuum and the relatively low spin states populated by the $^{3}$He reactions, $\gamma$-ray transitions of the lowest multipolarity are far more likely to occur than the higher ones. It is thus assumed that the main contribution to the experimental $\gamma$-ray transmission coefficient $\mathcal{T}(E_{\gamma})$ is of dipole character, i.e. $L = 1$. The $\gamma$-ray transmission coefficients are then essentially described by 
\begin{align}
\label{eq:exp_F}
 \mathcal{T}(E_{\gamma})&=\sum_{X L} \mathcal{T}_{X L}(E_{\gamma})\nonumber\\ 
&\approx  \left[ \mathcal{T}_{E 1}(E_{\gamma}) + \mathcal{T}_{M 1}(E_{\gamma})\right]~.
\end{align}
Under the assumption that there is an equal number of accessible states for both parities from any excitation energy and spin, the level density is expressed as
\begin{equation}
\label{eq:ld_eqnopar}
\rho(E_x,J,\pm\pi)=\frac{1}{2}\rho(E_x,J)~.
\end{equation}

\begin{figure}
\includegraphics[width=\columnwidth]{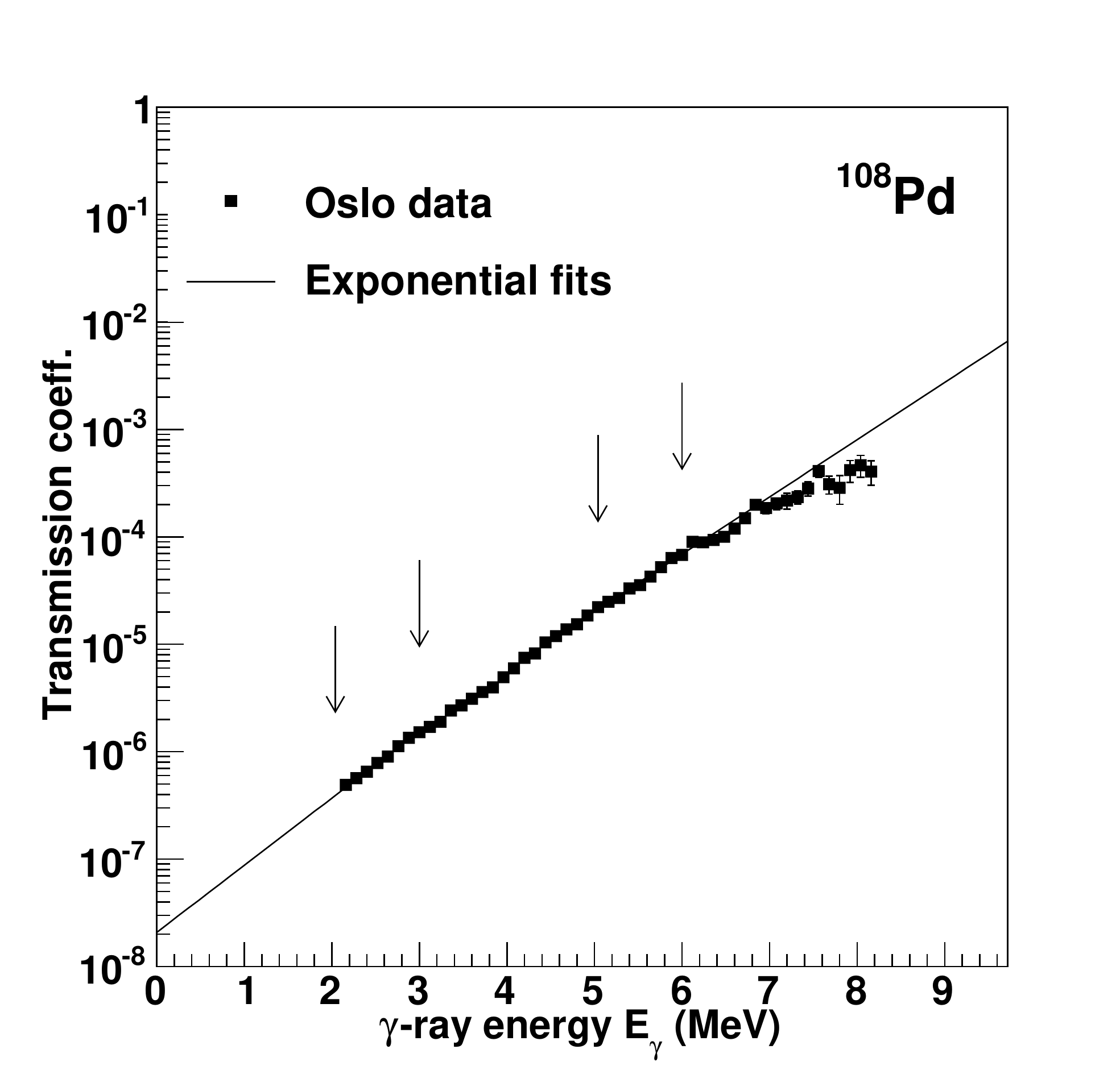}
\caption[T norm 108Pd]{Normalization of the transmission coefficients of $^{108}$Pd. The arrows indicate fit limits, and further details are provided in the text.}
\label{fig:T_norm}
\end{figure}

The average total radiative width of neutron capture resonances can be expressed in terms of the experimental $\gamma$-ray transmission coefficients as
\begin{align}
\label{eq:atrw_nr_F}
&\left< \Gamma_{\gamma}(B_n,J_f)\right>=\frac{B}{2\pi D_{0}}\int_{E_{\gamma}=0}^{B_n}dE_{\gamma} \mathcal{T}_{L=1}(E_{\gamma})\nonumber\\ 
&~~~~~~~\times \rho(B_n-E_{\gamma})\sum_{L^{\prime}=-1}^{1} g(B_n-E_{\gamma},J_f+L^{\prime})~,
\end{align}
where $B$ is the normalization coefficient. The spin distribution of the experimental level density is normalized so that $\sum_J g(E_x,J)\approx1$, for all available spins $J$. The experimental value of $\left< \Gamma_{\gamma}\right>$ at the neutron binding energy is then a weighted sum of the level widths of excited states with spin $J_f$, and the transformation coefficient $B$ can be determined by using the experimental $\left< \Gamma_{\gamma}(B_n)\right>$ and $D_{0}$ available in Ref.~\cite{mughabghab}.
Because of the integral in Eq.~(\ref{eq:atrw_nr_F}), the normalization requires transmission coefficients in the entire energy range $E_{\gamma}\in[0, B_n]$. The $\mathcal{T}(E_{\gamma})$ is therefore extrapolated with exponential functions at low and high energies. Normalization of the transmission coefficient for $^{108}$Pd is shown in Fig.~\ref{fig:T_norm}. Note that the high $E_{\gamma}$ exponential fit was performed to data somewhat lower than the highest points, which was due to achieve a better match to the normalization of the other isotopes.

\begin{figure}[h]
\includegraphics[width=\columnwidth]{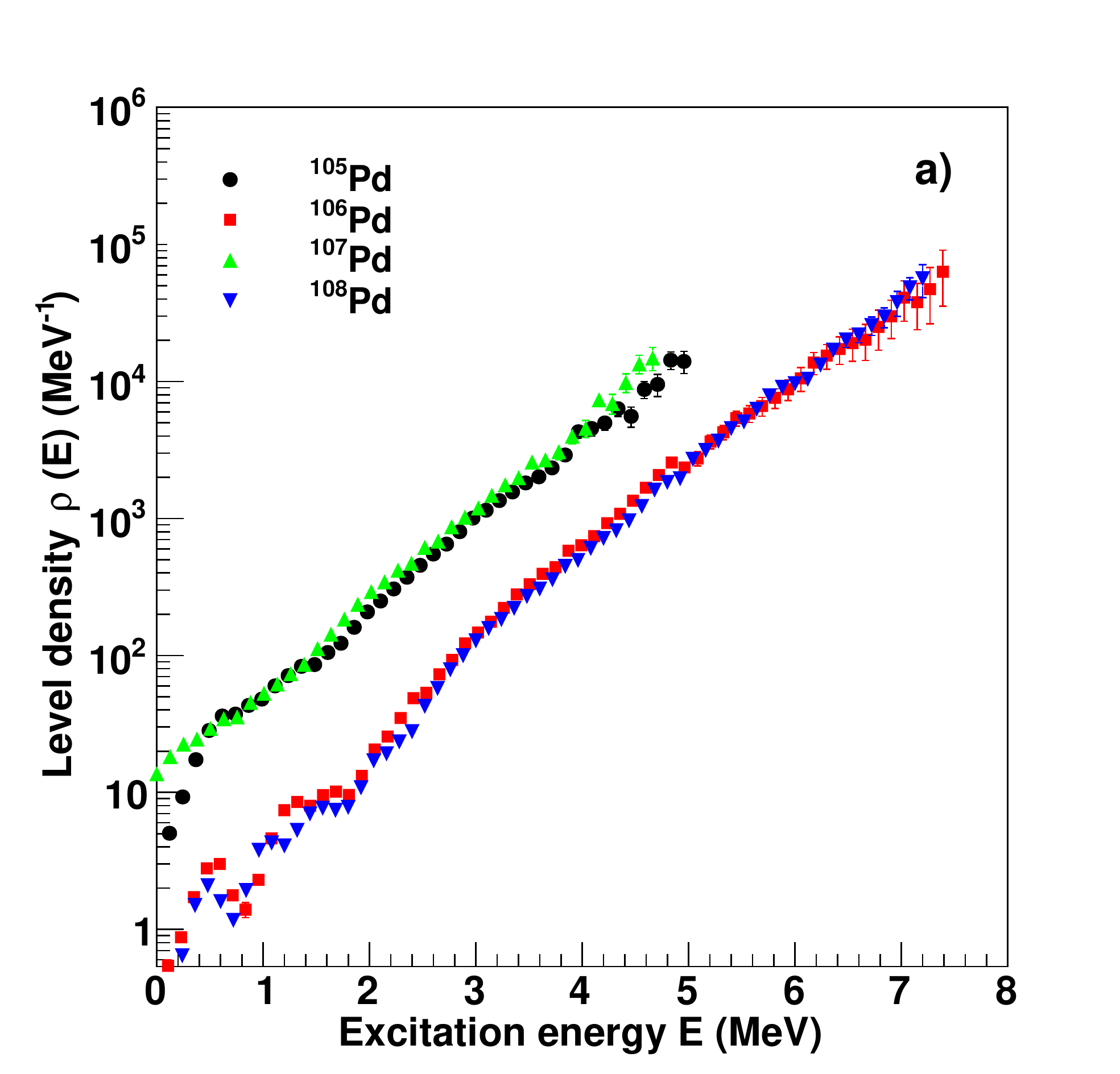}\\
\includegraphics[width=\columnwidth]{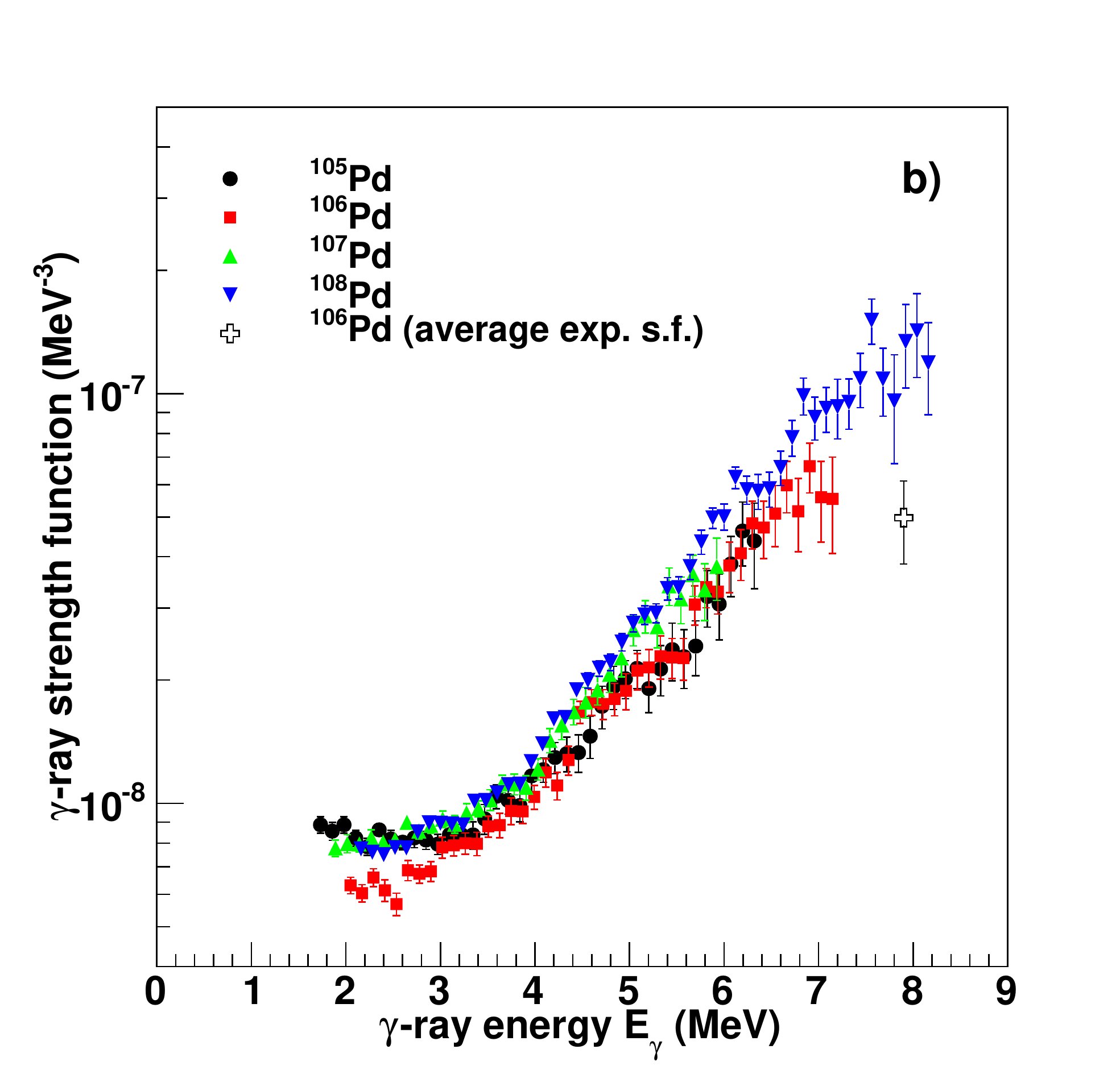}
\caption{(Color online) Extracted level densities (a) and $\gamma$-ray strength functions (b) of $^{105-108}$Pd.}
\label{fig:extfunc}
\end{figure}

Finally, the $\gamma$-ray strength function $f(E_{\gamma})$ is deduced through its relation to the $\gamma$-ray transmission coefficient~\cite{ripl2hb}
\begin{equation}
\label{eq:frelF}
\mathcal{T}_{X L}(E_{\gamma})=2\pi E_{\gamma}^{2L+1}f_{X L}(E_{\gamma})~,
\end{equation}
when assuming $L=1$ to be the dominating multipolarity for transitions in the quasi-continuum.
The parameters used for determining the normalization coefficients are provided in Tab.~\ref{tab:input_values}. The $\left< \Gamma_{\gamma} \right>$ value for $^{107}$Pd was not directly available for s-wave neutrons, but was deduced from two s-wave neutron resonances at higher energies listed in Ref.~\cite{mughabghab}. Note that for $^{108}$Pd the lowest value of $\left< \Gamma_{\gamma} \right>$ within the uncertainty had to be used in order to match the magnitude of the strength functions for the other isotopes. The extracted level densities and $\gamma$-ray strength functions, with recommended normalization, are depicted in Fig.~\ref{fig:extfunc}. The error bars represent statistical uncertainties and propagated errors from the unfolding and first-generation method, other uncertainties will be discussed later. The strength functions are also compared to the sum of the average $f_\textnormal{E1}$ and $f_\textnormal{M1}$ from~\cite{Av.SF}, which has slightly lower magnitude. However, it was not possible to obtain a normalization giving a lower magnitude for the strength functions and at the same time a good agreement between the level densities and strength functions for the measured isotopes. The normalization applied for the functions in Fig.~\ref{fig:extfunc} is therefore regarded as the best choice.

\begin{table*}
\caption[Input parameters]{Input parameters~\cite{nld_parameters, mughabghab} used in the normalization procedure.}
\begin{tabular}{ c | c  c  c  c  c | c  c  c | c }
\hline\hline
~Nucleus~ & ~$B_n$~ & ~$a$~ & ~$E_1$~ & ~$Pa'$~ & ~$\sigma(B_n)$~ & ~$I_t$~ & ~$D_{0}$~ & ~$\rho(B_n)$~ & ~$\left< \Gamma_{\gamma} \right>$~ \\
~ & ~[MeV]~ & ~[MeV$^{-1}$]~ & ~[MeV]~ & ~[MeV]~ & ~ & ~ & ~[eV]~ &  ~[$10^5$MeV$^{-1}$]~ & ~[meV]~ \\
\hline
~$^{105}$Pd~ & ~7.094~ & ~11.8~ & ~$-0.79$~ & ~0.199~ & ~$4.12^{+0.87}_{-0.41}$~ & ~0~ & ~$194(30)$~ & ~$1.80^{+1.30}_{-0.53}$~ & ~148(10)~\\
~$^{106}$Pd~ & ~9.561~ & ~12.8~ & ~$0.85$~ & ~2.625~ & ~$4.20^{+0.95}_{-0.42}$~ & ~$5/2$~ & ~$10.9(5)$~ & ~$7.09^{+3.11}_{-1.24}$~ & ~151(5)~\\
~$^{107}$Pd~ & ~6.536~ & ~12.5~ & ~$-0.73$~ & ~0.083~ & ~$4.09^{+0.84}_{-0.41}$~ & ~0~ & ~$174(25)$~ & ~$1.98^{+1.35}_{-0.57}$~ &~85(10)$^*$~\\
~$^{108}$Pd~ & ~9.228~ & ~13.4~ & ~$1.01$~ & ~2.613~ & ~$4.19^{+0.93}_{-0.42}$~ & ~$5/2$~ & ~$14.8(8)$~ & ~$5.20^{+2.31}_{-0.94}$~ & ~169(39)$^{**}$~\\
\hline\hline
\end{tabular}
\\$^*$Estimated from two s-wave resonances in Ref.~\cite{mughabghab}.
\\$^{**}$$\left< \Gamma_{\gamma} \right>=130$~meV is used in the normalization.
\label{tab:input_values}
\end{table*}

\section{Level densities}
\label{sec:ld}
The normalized level densities of the four nuclei are depicted in Fig.~\ref{fig:extfunc}a). They seem to be quite parallel above $\approx3$~MeV, which is satisfying since the level densities of neighboring nuclei are generally parallel on a logarithmic scale. The level densities are higher for the even-odd $^{105,107}$Pd isotopes due to the last valence neutron, which may occupy additional single-particle levels. This typically results in seven times the amount of accessible states for the even-odd nuclei above $\approx3$~MeV, as compared to their even-even $^{106,108}$Pd neighbor isotopes. Single particle levels are not accessible to even-even nuclei below the pair-breaking energy, and the energy required to break a nucleon pair in $^{106,108}$Pd is about $E_{\rm br}\approx2.8 - 2.9$~MeV. 
This energy is given by $E_{{\rm br},p(n)}\approx2\Delta_{p(n)}$, where the pair gap parameters $\Delta_{p(n)}$ are given by differences in binding energy $B_{p(n)}$, see e.g.~Ref.~\cite{pb_parameters}. The pair gap parameters for $^{105-108}$Pd are listed in Tab.~\ref{tab:pairbr}.
\begin{table}[b]
\caption{Nucleon pair gap parameters. See text for explanation.}
\begin{tabular}{ c | c | c }
\hline\hline
~Nucleus~ & ~$\Delta_n$~ & ~$\Delta_p$~\\
~ & ~[MeV]~ & ~[MeV]~\\
\hline
~$^{105}$Pd~ & ~$1.34$~ & ~$1.18$~\\
~$^{106}$Pd~ & ~$1.37$~ & ~$1.47$~\\
~$^{107}$Pd~ & ~$1.43$~ & ~$1.09$~\\
~$^{108}$Pd~ & ~$1.44$~ & ~$1.40$~\\
\hline\hline
\end{tabular}
\label{tab:pairbr}
\end{table}
The few excited states observed below the pair-breaking energy for $^{106,108}$Pd are caused by collective vibrational motion,
and breaking of pairs can be recognized in Fig.~\ref{fig:extfunc}a) as the level densities show a logarithmically constant increase in magnitude above the respective pair-breaking energies. 

Above the pair-breaking energies the characteristics of the level densities can be described by the constant temperature formula~\cite{nld_parameters} 
\begin{equation}
\label{eq:ct}
\rho_{\rm CT}(E_x)=\frac{1}{T} e^{(E_x-E_0)/T}~,
\end{equation}
where $T$ is the temperature, $E_x$ is the excitation energy and $E_0$ is the energy shift. As a test, the constant temperatures were estimated by letting $T$ and $E_0$ be free parameters, and fitting Eq.~(\ref{eq:ct}) to the level densities by a least square fit.
The fitted temperatures are provided in Tab.~\ref{tab:estT} and they seem to agree with each other, as well as with the predicted values of Ref.~\cite{nld_parameters} shown in column 3. The good agreement gives confidence to the slope found in the normalization procedure.
\begin{table}
\caption{Temperatures estimated from fitting the constant temperature formula, Eq.~(\ref{eq:ct}), to the experimental level densities. Column 1  indicates the fit limits. Column 3 shows temperatures from~\cite{nld_parameters}.}
\begin{tabular}{ c | c | c | c }
\hline\hline
~Nucleus~ & ~$E_{x,1} - E_{x,2}$~ & ~$T_{fit}$~ & ~$T_{\textnormal{CT}}$~\\
~ & ~[MeV]~ & ~[MeV]~ & ~[MeV]~\\
\hline
~$^{105}$Pd~ & ~$2.0-5.0$~ & ~$0.72^{+0.04}_{-0.05}$~ & ~$0.75(3)$~\\
~$^{106}$Pd~ & ~$3.0-6.0$~ & ~$0.71^{+0.02}_{-0.03}$~ & ~$0.71(2)$~ \\
~$^{107}$Pd~ & ~$2.0-4.7$~ & ~$0.69^{+0.04}_{-0.05}$~ & ~$0.70(4)$~ \\
~$^{108}$Pd~ & ~$3.0-6.0$~ & ~$0.68^{+0.02}_{-0.03}$~ & ~$0.68(2)$~ \\
\hline\hline
\end{tabular}
\label{tab:estT}
\end{table}

\section{Gamma-ray strength functions}
\label{sec:g-rsf}

The largest and most important resonances of atomic nuclei are the giant electric dipole resonance (GEDR) and the giant magnetic dipole resonance (GMDR). The GEDR accounts for most of the strength, and is often referred to as the Giant Dipole Resonance (GDR). 
In the following, the extracted $\gamma$-ray strength functions will be compared to empirical models developed for these resonances. These models are summarized in Ref.~\cite{ripl2hb}.
The models used in this work are the standard Lorentzian model for the magnetic dipole ($M1$) spin-flip resonance~\cite{bohrmott}, and the generalized Lorentzian model for the electric dipole ($E1$) resonance. The standard Lorentzian is described by~\cite{brink, paxel}
\begin{equation}
\label{eq:slo}
f_{M1}^{SLo}(E_{\gamma})=k\cdot\frac{\sigma_r E_{\gamma} \Gamma_r^2}{(E_{\gamma}^2-E_r^2)^2+E_{\gamma}^2\Gamma_r^2}~,
\end{equation}
where $\sigma_r$, $E_r$, and $\Gamma_r$ is the peak cross section, energy centroid, and width of the resonance respectively. The factor $k=(3 \pi^2 \hbar^2 c^2)^{-1}=8.674\cdot10^{-8}$~mb$^{-1}$MeV$^{-2}$ gives the conversion of the differential cross section (mb/MeV) to units of MeV$^{-3}$, which is the unit of the $\gamma$-ray strength function for dipole transitions. 
The generalized Lorentzian is described by~\cite{EGLo} 
\begin{align}
\label{eq:eglo}
f_{E1}^{GLo}&(E_{\gamma}, T)=\nonumber\\&k\cdot\sigma_r \Gamma_r \left[ \frac{E_{\gamma} \Gamma_{E_n}(E_{\gamma}, T)}{(E_{\gamma}^2-E_r^2)^2+E_{\gamma}^2\Gamma_{E_n}^2(E_{\gamma}, T)}\right. \nonumber\\ \nonumber\\ & ~~~~~~~~~~~~~~~~~~~~~~~~~~~\left.+  0.7\cdot\frac{\Gamma_{E_n}(0, T)}{E_r^3}\right]~,
\end{align}
where
\begin{equation}
        \Gamma_{E_n}(E_{\gamma}, T)=\frac{\Gamma_r}{E_r^2}(E_{\gamma}^2 + 4\pi^2 T^2)~.
\end{equation}
This model takes into account that the width of the $E1$ resonance is dependent of $E_{\gamma}$ and $T$, and includes a non-zero limit as $E_{\gamma}\rightarrow0$~MeV. 

The GDR models were fitted to experimental ($\gamma$,n) data from Ref.~\cite{utsunomiya}, and the resulting parameters are listed in Tab.~\ref{tab:modpar}. 
The temperatures, $T$, were taken as free, constant parameters in the fits. Further, the deformation parameter $\beta_2$ was also needed in order to calculate two-component GDR models, and this was taken from theoretically derived values calculated within the Finite Range Droplet Model (FRDM)~\cite{frdm95}.
\begin{table*}
\caption[Input parameters for model calculation]{The parameters used in the systematic GDR models.}
\begin{tabular}{ c | c  c  c | c  c  c | c  c  c | c | c }
\hline\hline
~Nucleus~ & ~$E_{r1}$~ & ~$\Gamma_{r1}$~ & ~$\sigma_{r1}$~ & ~$E_{r2}$~ & ~$\Gamma_{r2}$~ & ~$\sigma_{r2}$~ & ~$E_{r,M1}$~ & ~$\Gamma_{r,M1}$~ & ~$\sigma_{r,M1}$~ & ~$\beta_2$~ & ~$T$~ \\
~ & ~[MeV]~ & ~[MeV]~ & ~[mb]~ & ~[MeV]~ & ~[MeV]~ & ~[mb]~ & ~[MeV]~ & ~[MeV]~ & ~[mb]~ & ~ & ~[MeV]~ \\
\hline
~$^{105}$Pd~ & ~15.00~ & ~5.95~ & ~111.37~ & ~17.34~ & ~7.84~ & ~55.68~ & ~8.69~ & ~4.0~ & ~1.35~ & ~0.171~ & ~$0.50^{+0.17}_{-0.05}$\\
~$^{106}$Pd~ & ~15.06~ & ~5.99~ & ~111.41~ & ~17.40~ & ~7.90~ & ~55.71~ & ~8.66~ & ~4.0~ & ~1.26~ & ~0.171~ & ~$0.47^{+0.08}_{-0.15}$~\\
~$^{107}$Pd~ & ~14.79~ & ~5.79~ & ~113.12~ & ~17.48~ & ~7.97~ & ~56.56~ & ~8.64~ & ~4.0~ & ~1.37~ & ~0.198~ & ~$0.51^{+0.17}_{-0.04}$~\\
~$^{108}$Pd~ & ~14.53~ & ~5.60~ & ~118.70~ & ~17.07~ & ~7.61~ & ~59.35~ & ~8.61~ & ~4.0~ & ~1.43~ & ~0.190~ & ~$0.49^{+0.27}_{-0.08}$~\\
\hline\hline
\end{tabular}
\label{tab:modpar}
\end{table*}

\begin{figure*}
\includegraphics[width=\textwidth]{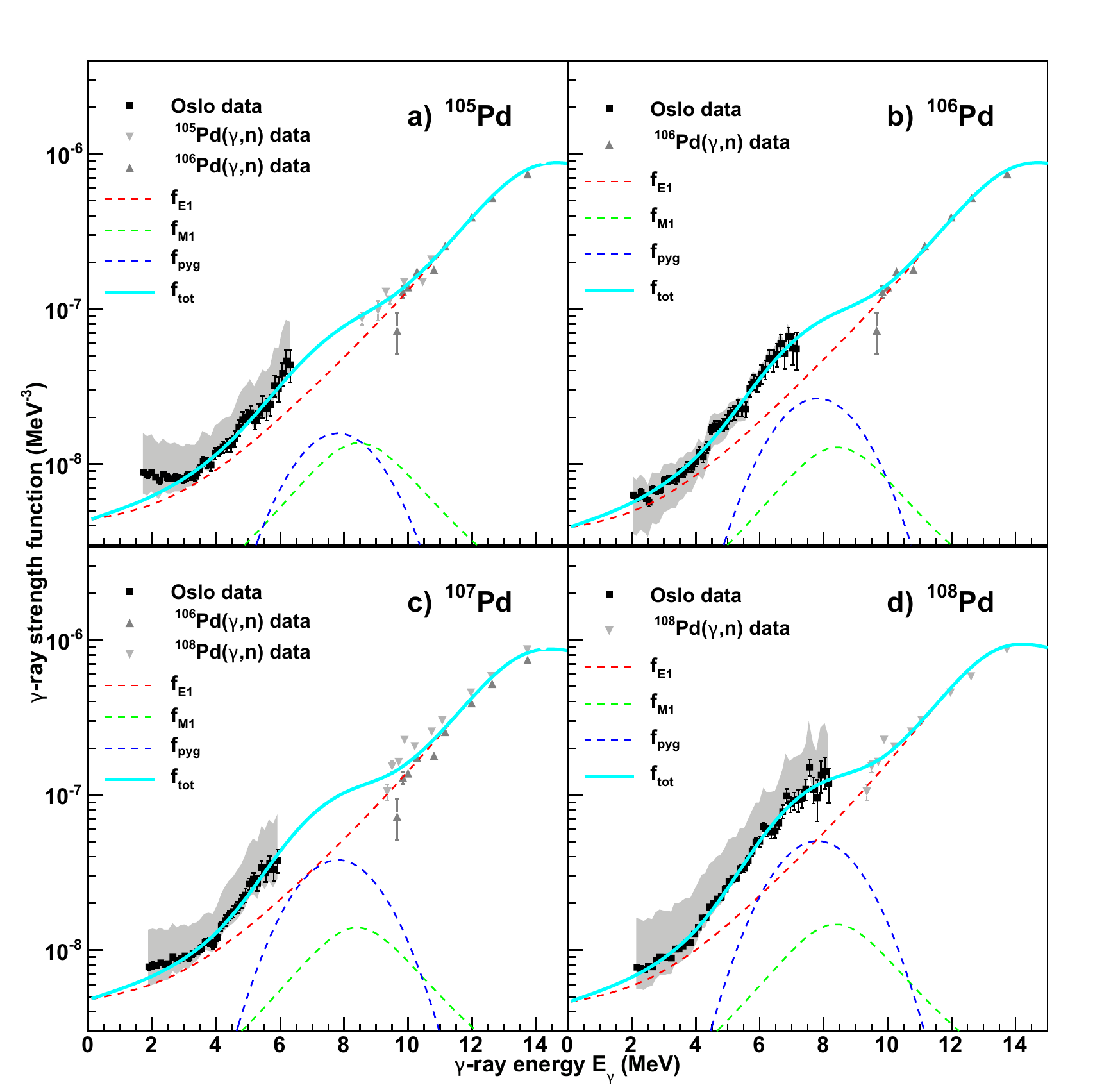}
\caption{(Color online) The extracted $f(E_\gamma)$ compared to models. The shaded area indicates systematical errors in the normalization procedure due to uncertainties in $\sigma$, $D_0$, and $\left< \Gamma_\gamma \right>$. The error bars of the Oslo data contain statistical errors, as well as uncertainties in the unfolding and extraction of first-generation $\gamma$-ray spectra.}
\label{fig:strengthfit}
\end{figure*}

Figures~\ref{fig:strengthfit}a) - \ref{fig:strengthfit}d) depict the extracted $\gamma$-ray strength functions, experimental ($\gamma$,n) data, and giant dipole resonance models for comparison. The PDR model, which is also included in these figures, will be explained in the following discussion.

First of all, an abrupt enhancement is observed in the experimental $\gamma$-ray strength functions for $E_{\gamma}>4$~MeV, as compared to the GDR. This feature is interpreted as a PDR, and seems to be increasing as a function of neutron number. Enhancement for $E_{\gamma}<4$~MeV is evident for $^{105}$Pd, but not for the other isotopes. The observations correspond to discoveries made for cadmium isotopes~\cite{Cd-isotopes}, which showed a resonance for $E_{\gamma}>4$~MeV, and also enhanced strength at low energy, i.e.~$E_{\gamma}<4$~MeV. 
The results showed that both the enhancements above and below $E_\gamma\approx4$~MeV seems to be dependent on neutron number, with the low-energy enhancement inversely proportional, and the PDR strength proportional to the number of neutrons. However, it should be stressed that it is not believed to be any connection between the two enhancement mechanisms; the limit of $E_{\gamma}=4$~MeV is simply used as a delimiter to distinguish them.    
Further, the enhancement in the $\gamma$-ray strength functions above $E_\gamma\approx4$~MeV is also very similar to the results found for tin isotopes~\cite{tin_pyg1, tin_pyg2}. However, a neutron number dependency of the strength was not observed for these isotopes, and the tin isotopes also completely lacked the low-energy enhancement. 

As for the Sn- and Cd isotopes, it was not possible to make a good fit to the pygmy resonance with a single Lorentzian distribution Eq.~(\ref{eq:slo}). When this problem was encountered for the Sn- and Cd isotopes, a single Gaussian distribution was used instead. Such a Gaussian shape has also been used for the E1 pygmy in exotic, neutron-rich nuclei, e.g.~for $^{68}$Ni~\cite{gauss68Ni}. It is possible that this single Gaussian distribution represents the sum of a number of narrow Lorentzians, but there is no theoretical foundation to our knowledge that supports this. In order to keep the fit on a basic level in terms of free parameters, the pygmy resonance was chosen to be fitted by a single Gaussian distribution also in this work,
\begin{equation}
\label{eq:pygmy}
f_{\rm pyg}=k\cdot \sqrt{\frac{2}{\pi}} \cdot \frac{\sigma_{\rm pyg}}{\Gamma_{\rm pyg}} e^{-2(E_{\gamma}-E_{\rm pyg})^2/\Gamma_{\rm pyg}^2}~,
\end{equation}
where the functional form has been expressed in such a way that the PDR parameters follow the same notation as the GDR parameters.

A systematic investigation of the $\gamma$-ray strength functions was performed by adopting the following description of the total strength
\begin{equation}
\label{eq:str_tot_c}
f_{\rm tot}=f_{E1}^{GLo} + f_{M1}^{SLo} + f_{\rm pyg}~,
\end{equation}
and fitting it to the experimental data. Note that the $f_{E1}$ and $f_{M1}$ resonance parameters were maintained, and the parameters $E_{\rm pyg}$ and $\Gamma_{\rm pyg}$ were adopted from a free fit to the $^{108}$Pd data. This was done because the dataset of $^{108}$Pd covers the largest range, and because it is assumed that the centroid and width of the PDR do not change considerably for the neighboring nuclei.
Thus, only the pygmy resonance parameter $\sigma_{\rm pyg}$ was treated as a free parameter for the $^{105-107}$Pd isotopes.
The parameters were determined by a least square fit, and the resulting values are shown in Tab.~\ref{tab:fit_pyg}. It is possible to identify systematic trends among the peak cross sections, i.e.~they seem to increase by a fixed value ($\approx0.5$~mb) with every additional neutron.  

\begin{table}
\caption{Fitted parameters of the pygmy resonances.} 
\begin{tabular}{ c | c  c  c }
\hline\hline
~Nucleus~ & ~$E_{\rm pyg}$~ & ~$\Gamma_{\rm pyg}$~ & ~$\sigma_{\rm pyg}$~  \\
~ & ~[MeV]~ & ~[MeV]~ & ~[mb]~  \\
\hline
~$^{105}$Pd~ & ~$7.81$~ & ~$2.81$~ & ~$0.64^{+0.76}_{-0.40}$~  \\
~$^{106}$Pd~ & ~$7.81$~ & ~$2.81$~ & ~$1.08^{+0.47}_{-0.16}$~ \\
~$^{107}$Pd~ & ~$7.81$~ & ~$2.81$~ & ~$1.55^{+2.59}_{-0.89}$~  \\
~$^{108}$Pd~ & ~$7.81$~ & ~$2.81$~ & ~$2.05^{+3.14}_{-0.26}$~ \\
\hline\hline
\end{tabular}
\label{tab:fit_pyg}
\end{table}

As can be seen in Figs.~\ref{fig:strengthfit}a) - d) the pygmy resonance is well reproduced by a Gaussian distribution.
The shaded area in the figures represents errors imposed by uncertainties in the spin cut-off parameter $\sigma$, neutron resonance parameter $D_0$, and the radiative width $\left< \Gamma_{\gamma} \right>$. It is assumed that neighboring isotopes have more or less overlapping strength functions, and the errors are thus constrained by the small uncertainties of $^{106}$Pd data.

Assuming that all the PDR strength is caused by $E1$ transitions, the integrated pygmy strengths were compared to the TRK sum rule~\cite{TRK1, TRK2, TRK3} 
\begin{equation}
\label{eq:TRKsr}
\sigma_{\rm TRK}\approx 60\frac{NZ}{A}{\textnormal{~MeV~mb}}. 
\end{equation}
In Tab.~\ref{tab:int_str} it can be seen that the ratio of the integrals increase by $\approx0.3$~\% with increasing neutron number. 

\begin{table}
\caption{Integrated strengths of the pygmy resonances.}
\begin{tabular}{ c | c  c  c }
\hline\hline
~Nucleus~ &  ~$\sigma_{\rm TRK}$~ & ~$\sigma_{\rm pyg,int}$~ &  ~$\sigma_{\rm pyg,int}/\sigma_{\rm TRK}$~ \\
~ &  ~[MeV~mb]~ & ~[MeV~mb]~ &  ~\%~ \\
\hline
~$^{105}$Pd~ & ~$1550.86$~ & ~$6.41^{+7.59}_{-4.01}$~ & ~$0.41^{+0.49}_{-0.26}$~ \\
~$^{106}$Pd~ & ~$1562.26$~ & ~$10.76^{+4.74}_{-1.56}$~ & ~$0.69^{+0.30}_{-0.10}$~ \\
~$^{107}$Pd~ & ~$1573.46$~ & ~$15.47^{+25.93}_{-8.87}$~ & ~$0.98^{+1.65}_{-0.56}$~ \\
~$^{108}$Pd~ & ~$1584.44$~ & ~$20.52^{+31.38}_{-2.62}$~ & ~$1.30^{+1.98}_{-0.17}$~ \\
\hline\hline
\end{tabular}
\label{tab:int_str}
\end{table}

\begin{figure}
\includegraphics[width=\columnwidth]{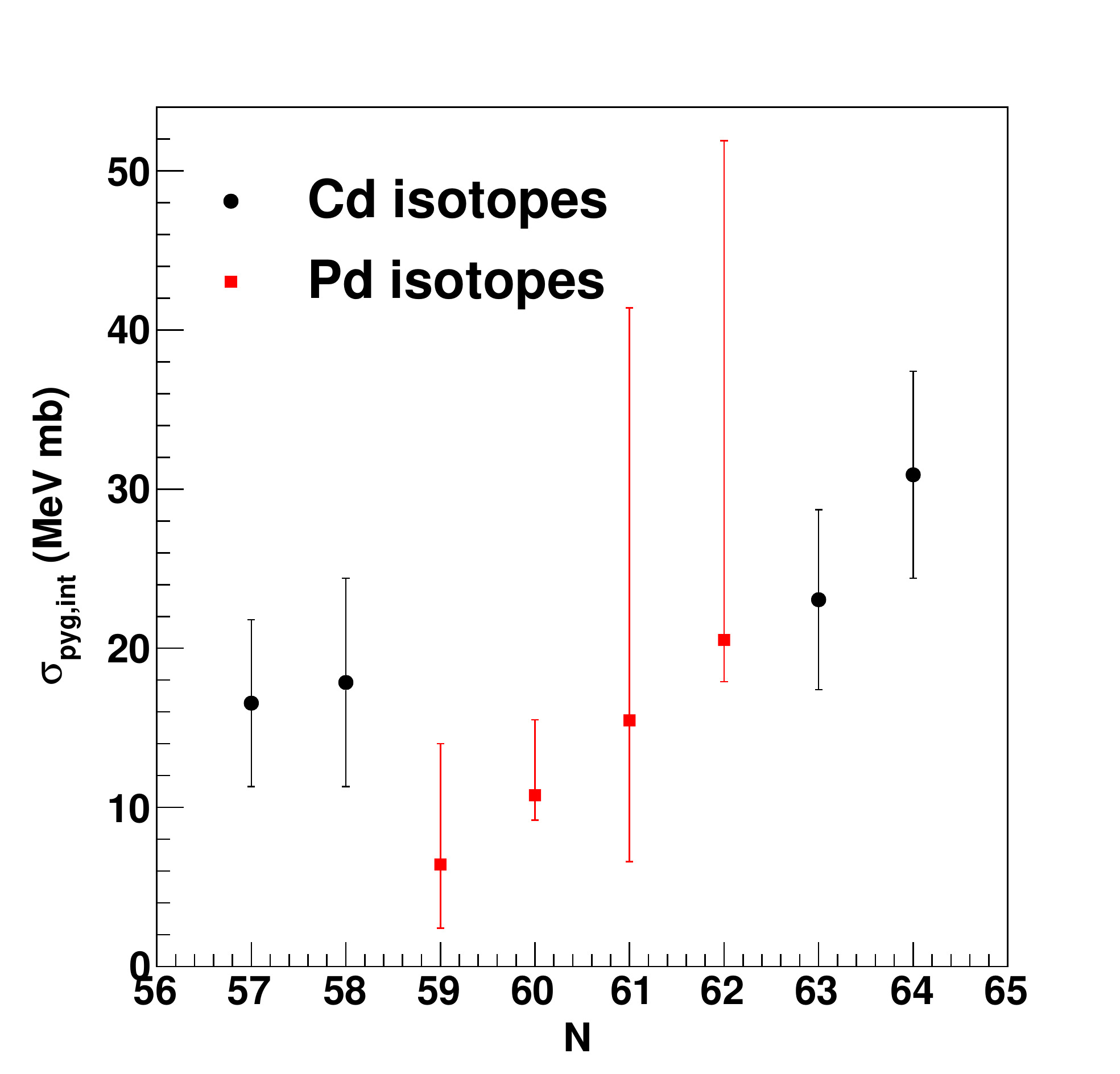}
\caption{(Color online) The integrated PDR strengths of the Cd~\cite{Cd-isotopes} and Pd isotopes.}
\label{fig:intstrcomp}
\end{figure}

A commonly accepted explanation for the PDR is that a collective skin of excess neutrons oscillate with respect to a $Z\approx N$ core~\cite{nskin}. Microscopic calculations have been performed within this picture, with promising results~\cite{nskinmic}. However, the collectivity of the resonance is still under debate, and another set of microscopic calculations~\cite{nonskin} actually oppose the idea of a collective mode. 
The latter work states that the resonance might instead be caused by rapidly varying particle-hole excitations, which are said to be mixed proton and neutron excitations, and that the neutrons carry more strength. Both the collective- and non-collective pictures might thus explain the neutron number dependency of the PDR strength. Figure \ref{fig:intstrcomp} shows the integrated PDR strengths of the cadmium and palladium isotopes plotted as a function of neutron number. 
The data indicate that the PDR strength increases as a function of neutron number, however, the functional form cannot be determined due to the large error bars.

\section{Summary and conclusions}
\label{sec:summary}

The level densities and $\gamma$-ray strength functions of $^{105-108}$Pd have been extracted and analyzed. The recommended normalization of the extracted data is supported by the good agreement between all the data sets.

The level densities seem to correspond well to known characteristics. The temperatures deduced from the logarithmically constant slopes of the level densitites agree very well in value, both compared to each other and to empirical values. This indicates that the level densities are quite parallel, and further supports the slopes determined in the normalization procedure. The extracted level densities may be used for further investigation of the thermodynamic properties of the Pd isotopes.

The $\gamma$-ray strength functions were compared to parameterized GDR models, and for $E_{\gamma}>4$~MeV they all exhibited an abrupt enhancement of the strength relative to these models. This corresponds to previous observations for tin- and cadmium isotopes. The $^{105}$Pd data also clearly indicate a low-energy enhancement, in contrast to the other Pd isotopes which have only very weak or no indications of this. These findings are consistent with previous observations for the cadmium isotopes, and supports the idea of a transitional region.

There are rather large uncertainties in the deduced PDR data, but however, when assuming that the $\gamma$-ray strength functions should be very similar for neighboring nuclei, the most reasonable values are constrained by the low uncertainty of the $^{106}$Pd data. 
The results show that the strength of the pygmy resonance increases as a function of neutron number, which indicates that the resonance is related to the excess neutrons in a systematical way. The nature of the resonance can not be concluded based on this behavior, because both the collective and non-collective pictures suggest a neutron dependency. In the collective neutron-skin picture it is trivial that the strength increases with excess neutrons, and in the non-collective picture it is stated that most of the strength is carried by neutrons. 
However, the Gaussian shape of the PDR suggests that there is a large number of narrow resonances in this energy region, which opposes the idea of a single collective neutron-skin resonance.

\begin{acknowledgements}
The authors would like to thank the University of Cologne for providing the high quality Pd target.
The authors also gratefully acknowledges funding from the Research Council of Norway, project grant no. 210007. At last, we would like to thank the operators E. A. Olsen, A. Semchenkov, and J. Wikne for providing optimal experimental conditions.
\end{acknowledgements}




\end{document}